\documentclass[english,12pt]{article}
\usepackage[]{fontenc}
\usepackage{graphicx}
\usepackage[authoryear]{natbib}

\makeatletter

\date{}
\usepackage{hyperref}
\def\E{{\bf{\mathcal{E}}}}
\def\B{{\bf{\mathcal{B}}}}
\def\spin{{\bf{\sigma}}}

\newcommand{\W}{\mathcal{W}}
\newcommand{\tbar}{\bar{t}}
\newcommand{\rbar}{\bar{r}}
\def\phivector{\mbox{\boldmath${\phi}$}}

\makeatother

\usepackage{babel}

\begin{document}

\title{Spin-String Interaction in QCD Strings}

\author{Vikram Vyas\\
Physics Department \\
 St. Stephen's College, Delhi University, Delhi\\
vvyas@stephens.du.ac.in}

\maketitle
\begin{abstract}
I consider the question of the interaction between a QCD string and
the spin of a quark or an antiquark on whose worldline the string
terminates. The problem is analysed from the point of view of a string
representation for the expectation value of a Wilson loop for a spin-half
particle. A string representation of the super Wilson loop is obtained
starting from an effective string representation of a Wilson Loop.
The action obtained in this manner is invariant under a worldline
supersymmetry and has a boundary term which contains the spin-string
interaction. For rectangular loops the spin-string interaction vanishes
and there is no spin-spin term in the resulting heavy quark potential.
On the other hand if an allowance is made for the finite intrinsic
thickness of the flux-tube, by assuming that the spin-string interaction
takes place not just at the boundary of the string world-sheet but
extends to a distance of the order of the intrinsic thickness of the
flux tube, then we do obtain a spin-spin interaction which falls as
the fifth power of the distance. Such a term was previously suggested
by Kogut and Parisi in the context of a flux-tube model of confinement. 
\end{abstract}

\section{Introduction}

There is strong numerical evidence that a flux-tube is formed between
a static quark and an anti-quark when the separation between them
is of the order of a fermi or even less, and that such flux-tubes
can be described by effective string models (for a review see for
e.g \citet{Bali:2001aa,Kuti:2005xg} ). This evidence for the formation
of a flux-tube and its string like behaviour matches well with the
fact that the spectrum of highly excited mesons are well described
by open-string models of mesons. Further these facts are in concordance
with the idea that in a suitable limit, namely in the limit of a large
number of colors, QCD is exactly equivalent to some unknown fundamental
string theory (for a contemporary review of these idea see for e.g
\citet{Brower:2005cd}.) It is therefore natural to ask a more detailed
question about the dynamics of the QCD string, namely, do the spin
of the quark and the anti-quark interact with the string connecting
them? Such an interaction could lead to a long range spin-spin term
in the heavy quark potential \citet{Kogut:1981aa}. Spin-string interaction
could also perhaps be responsible for the pion-rho mass difference
in effective string models of meson \citet{Polchinski:1992vg}. More
generally the spin-string interaction could help answer the question
of how is spontaneous breaking of chiral symmetry reflected in a fundamental
string representation of QCD.

The nature of the interaction between the spin of the quark and the
string has been investigated in the context of open string models
of mesons (see for e.g. \citet{Pisarski:1987jb,Pawelczyk:1993aa,Allen:2003kd}.)
In the present investigation we will take a different approach. We
will start with the assumption that the expectation value of the Wilson
loop over the gauge fields can be written as a sum over surfaces whose
boundary is the given loop \citet{Wilson:1974aa,Nambu:aa,Luscher:1980fr,Luscher:1980ac,Polyakov:1987ez}.
These surfaces can be regarded as the worldsheets of a string whose
end points lie on the loop, while the loop represents the worldline
of a scalar particle-antiparticle pair that is created at a point
and is annihilated latter. If we replace the closed worldline of a
scalar particle by a closed worldline of a spin-half particle then
the amplitude for the corresponding process is given, apart from the
kinematic factors, by the Wilson loop for a spin-half particle \citet{Feynman:1951gn,Berezin:1976eg,Brink:1976sz,Brink:1976uf,Barducci:1976qu}.
Such a Wilson loop is often referred to as a super Wilson loop as
it is invariant under a one-dimensional supersymmetry\citet{Migdal:1998aa}.
If we can write the expectation value of a super Wilson loop as a
sum over the surface whose boundary is the given loop, then the corresponding
string action automatically includes the spin of the quark and the
spin-string interaction %
\footnote{This approach was previously considered in Ref.\citet{Vyas:2000qi}
within the context of Polyakov's confining strings \citet{Polyakov:1996nc}.%
}. The task of finding the string representation of a super Wilson
loop is facilitated by the fact that the super Wilson loop is not
an independent loop functional but is related to the Wilson loop via
the area derivative of a loop \citet{Migdal:1998aa}.

The simplest string action used to model QCD strings is the Nambu-Goto
action which is the area of the string world-sheet. Though the Nambu-Goto
string in four dimensions suffers from serious problems, but it can
be thought of as the leading term in an effective description \citet{Polchinski:1991ax,Drummond:2004yp,Hari-Dass:2006aa,Hari-Dass:2006ab}.
The success of Nambu-Goto string in modelling the heavy quark potential
as obtained from the lattice QCD simulations \citet{Luscher:2002aa,Caselle:2005xy,HariDass:2006pq}
indicates that the expectation value of the Wilson loop over the gauge
fields can be well represented by a sum over surfaces with the surface
being weighted by the exponential of the Nambu-Goto action, at least
for rectangular loops. With this as our justification, we will obtain
a string representation for the expectation value of the super Wilson
loop via the area derivative of the Nambu-Goto action. The super Wilson
loop, when written in terms of anticommuting variables, is invariant
under a worldline supersymmetry. We will verify that the action of
the string representing the super Wilson loop is also invariant under
the worldline SUSY.

In the string representation of the super Wilson loop the spin-string
interaction appears as a boundary term, representing interaction between
the spin of the quark (or the antiquark) and the extrinsic curvature
of the worldsheet at the boundary. To obtain some intuition about
the significance of the spin-string interaction we calculate the expectation
value of a rectangular super Wilson loop, from which one can extract
the spin-dependent heavy quark potential \citet{Eichten:1980mw,Peskin:1983aa,Gromes:1984ma,Barchielli:1986zs,Barchielli:1988zp,Brambilla:1993zw,Pineda:2000sz,Brambilla:2004jw}.
It turns out that for a rectangular super Wilson loop the spin-string
term vanishes, and therefore there is no spin-spin dependent term
in the heavy quark potential.

But if we think of a string as an effective description for a flux-tube
of finite \emph{intrinsic} width that is formed between a static quark-antiquark
pair, and evaluate the spin-string interaction not right at the boundary
of the rectangular loop but average it over a distance of the order
of the thickness of the flux-tube, then we do obtain a spin-spin interaction
term. The form of this term is precisely the one considered by Kogut
and Parisi in the context of a fluctuating flux-tube model of confinement
\citet{Kogut:1981aa}. This term represents an attractive interaction
between anti-aligned spins which falls as the fifth power of the inverse
distance between the quark and the antiquark.

The outline of the paper is the following: in the next section the
physical significance of the Wilson loop and the super Wilson loop
are recalled and their relationship via area derivative of the loop
is stated. A string representation of the super Wilson loop is obtained
in sec.(\ref{sec:introSWL}), assuming that the string representation
of the Wilson loop is provided by the Nambu-Goto action. It is also
shown that the string action for the super Wilson loop is invariant
under the worldline SUSY, and a brief comment on the relationship
between the string representation of a super Wilson loop and the vacuum
expectation value of chiral condensate is made. In sec.(\ref{sec:SWLstaticPot})
the string representation of the super Wilson loop is used to obtain
the expectation value of a rectangular super Wilson loop from which
the heavy quark potential is obtained. It is found that the spin-string
interaction vanishes and therefore there is no spin-spin dependent
correction to the heavy quark potential. Next, in sec.(\ref{sec:fluxTubeStaticPot})
we evaluate the spin-string interaction in the spirit of the flux-tube
model and obtain a non-vanishing spin-spin interaction. The conclusions
are stated in the final section.

\section{The Wilson Loop and the Super Wilson Loop\label{sec:introSWL}}

The Wilson loop for a scalar particle in the fundamental representation
of the gauge group is defined as, \begin{equation}
W[x(\tau)]={\rm Tr}\hat{P}\exp\left\{ i\oint d\tau A.\frac{dx}{d\tau}\right\} ,\label{eq:WL}\end{equation}
 where the trace is over the color indices of the matrix valued vector
potential, $A=A^{a}.\tau_{a}$, with $\tau_{a}$ as the matrices providing
the fundamental representation of the Lie-algebra of the gauge group.
$\hat{P}$ is the path ordering operator that instructs us to order
the color matrices along the loop $x(\tau)$ in the order of the increasing
value of the parameter $\tau$. To recall the physical significance
of the Wilson loop \citet{Wilson:1974aa}, consider the propagation
of a meson which is created at $x_{i}$ and annihilated at $x_{f}$
in the approximation in which one neglects the virtual quark pairs.
In this approximation the amplitude for this process can be written
as a sum over closed paths passing through points $x_{i}$ and $x_{f}$,
each path being weighted by the expectation value of the corresponding
Wilson loop and some kinematic factors. The expectation value of the
Wilson loop being defined as

\begin{eqnarray}
<W[x(\tau)]>_{YM} & = & \frac{1}{Z_{YM}}\int DA\exp(-S_{YM}[A])W[x(\tau)],\label{eq:Wym}\\
Z_{YM} & = & \int DA\exp(-S_{YM}[A]),\label{eq:Zym}\end{eqnarray}
 where $S_{YM}[A]$ is the Euclidean action for the gauge fields.
One way of formulating gauge-string duality is to assume that the
expectation value of the Wilson loop can be written as a sum over
surfaces, \begin{equation}
<W[x(\tau)]>_{YM}=\int DX\exp\left(-S_{WL}[X]\right),\label{eq:Wstring}\end{equation}
 where $X(\sigma)$ is the surface whose boundary is the loop $x(\tau)$
and $S_{WL}[X]$ is some unknown string action \citet{Wilson:1974aa,Nambu:aa,Luscher:1980fr,Luscher:1980ac,Polyakov:1987ez}.

In the above discussion the particle was assumed to be a scalar particle,
if we want to describe the propagation of a meson, including the spin
of the quark and the antiquark, then the role of the Wilson loop is
played by the Wilson loop for a spin-half particle \citet{Feynman:1951gn}
(see \citet{Peskin:1983aa} for a review)\begin{eqnarray}
\W[x(\tau),\gamma_{\mu}(\tau)] & = & {\rm Tr}\hat{P}\exp\left\{ i\oint d\tau\frac{dx}{d\tau}.A-\frac{i}{4}\oint{d\tau\gamma}_{\mu}\gamma_{\nu}F_{\mu\nu}\right\} \label{eq:WLspinHalf}\\
 & = & {\rm Tr}\hat{P}\exp\left\{ i\oint dt\dot{x}\cdot A+\frac{1}{4}\oint{d\tau\Sigma}_{\mu\nu}F_{\mu\nu}\right\} \end{eqnarray}
 where $\gamma_{\mu}$ are the Dirac gamma matrices and $\Sigma_{\mu\nu}$
are the corresponding spin matrices. Since these matrices do not commute
therefore they too have to be path-ordered and in that sense they
are function of the loop parameter $\tau$. In the context of the
path integral for a spin-half particle the appropriate Wilson loop
can also be written using Grassmann variables,\begin{equation}
\W[x(\tau),\psi(\tau)]={\rm Tr}\hat{P}\exp\left\{ i\oint d\tau\left(\frac{dx}{d\tau}.A-\frac{1}{2}\psi_{\mu}\psi_{\nu}F_{\mu\nu}\right)\right\} ,\label{eq:SWL}\end{equation}
 where $\psi(\tau)$ are four independent anti-commuting variables
\citet{Berezin:1976eg,Brink:1976sz,Brink:1976uf,Barducci:1976qu}.
Their role is same as that of gamma matrices, the integration over
$\psi(\tau)$ with suitable action for a free spin-half particle is
equivalent to taking trace over the gamma matrices. An immediate advantage
of writing the Wilson loop for a spin-half particle using $\psi(\tau)$
is that it is invariant under the following one-dimensional supersymmetry\begin{equation}
\delta x=\epsilon\psi;\,\delta\psi=-\epsilon\dot{x}.\label{eq:wlineSUSY}\end{equation}
 For this reason, in what follows we will refer both to (\ref{eq:WLspinHalf})
and to (\ref{eq:SWL}) as the super Wilson Loop. The super Wilson
loop is not an independent loop functional but is related via a linear
operator to the Wilson loop \begin{equation}
\exp\left\{ -\frac{1}{2}\oint d\tau\psi_{\mu}\psi_{\nu}\frac{\delta}{\delta\sigma_{\mu\nu}}\right\} W[x(\tau)]=\W[x(\tau),\psi(\tau)],\label{eq:WloopToSWL}\end{equation}
 where $\frac{\delta}{\delta\sigma_{\mu\nu}}$ is the area derivative
of the loop \citet{MIGDAL:1983aa,Migdal:1998aa}.

\section{String Representation of Super Wilson Loop\label{sec:stringSWL}}

The quark-antiquark potential is surprisingly well modelled by a Nambu-Goto
string\citet{Luscher:2002aa,Caselle:2005xy,HariDass:2006pq}, suggesting
that at least for rectangular loops the expectation value of the Wilson
loop can be written as, \begin{equation}
<W[x(\tau)]>_{YM}=\int DX(\sigma)\exp\left\{ -S_{NG}[X(\sigma)]\right\} .\label{eq:NGstringDual}\end{equation}
 The Nambu-Goto action, $S_{NG}$, is given by \begin{equation}
S_{NG}[X(\sigma)]=T_{0}\int d^{2}\sigma\sqrt{g},\label{eq:NGaction}\end{equation}
 where $T_{0}$ is the string tension and $g$ is the determinant
of the induced metric. The induced metric can be written using the
world-sheet coordinates, $(\sigma_{1},\sigma_{2})$, as\begin{equation}
g_{ab}[\sigma]=\frac{\partial X}{\partial\sigma_{a}}\cdot\frac{\partial X}{\partial\sigma_{b}}.\label{eq:g}\end{equation}
 Using Eq. (\ref{eq:WloopToSWL}) one can write the expectation value
of the super Wilson loop in terms of the expectation value of the
Wilson loop,\begin{eqnarray}
<\W[x(\tau),\psi(\tau)]>_{YM} & = & <\exp\left\{ -\frac{1}{2}\oint d\tau\psi_{\mu}\psi_{\nu}\frac{\delta}{\delta\sigma_{\mu\nu}}\right\} W>_{YM},\nonumber \\
 & = & \exp\left\{ i\oint d\tau\psi_{\mu}\psi_{\nu}\frac{\delta}{\delta\sigma_{\mu\nu}}\right\} \int DX\exp\left\{ -S_{NG}[X]\right\} ,\nonumber \\
 & = & \int DX\exp\left\{ -S_{SWL}[X,x(\tau),\psi(\tau)]\right\} ,\label{eq:defStringSWL}\end{eqnarray}
 where the string action for the super Wilson loop is \begin{equation}
S_{SWL}=T_{0}\int d^{2}\sigma\sqrt{g}-\frac{T_{0}}{2}\oint d\tau\psi_{\mu}(\tau)\psi_{\nu}(\tau)t_{\mu\nu}(\tau).\label{eq:SNambuGoto}\end{equation}
 In obtaining the above action we have used the fact \citet{MIGDAL:1981aa}
that the area derivative of the area functional is \begin{equation}
\frac{\delta}{\delta\sigma_{\mu\nu}(\sigma(x(\tau))}\int d^{2}\sigma'\sqrt{g}=t_{\mu\nu}[\sigma(x(\tau))],\label{eq:areaDerivativeNG}\end{equation}
 and $t_{\mu\nu}$ is given by \begin{equation}
t_{\mu\nu}(\sigma)=\frac{\epsilon^{ab}\partial_{a}X_{\mu}\partial_{b}X_{\nu}}{\sqrt{g}}=\frac{X_{\mu\nu}(\sigma)}{\sqrt{g}}.\label{eq:defTmunu}\end{equation}
 Thus, the action for the super Wilson loop loop differ from the Nambu-Goto
action by the presence of an additional boundary term. The boundary
term represents the interaction between the string variables and the
spin of the quark whose worldline is the boundary of the given loop. 

As mentioned earlier, super Wilson loop is invariant under a worldline
SUSY (\ref{eq:wlineSUSY}), which we will refer to as SUSY1. One expects
that the action (14) too should be invariant under SUSY1 %
\footnote{I would like to thank V. P. Nair for emphasising this to me.%
}. We can check this using the methods of loop calculus. \citet{MIGDAL:1983aa,Makeenko:1999hq}.
To do so, let us write the action (\ref{eq:SNambuGoto}) as$ $ \begin{eqnarray}
S_{SWL} & = & T_{0}\int d^{2}\sigma\sqrt{g}-\frac{T_{0}}{2}\oint d\tau\psi_{\mu}(\tau)\psi_{\nu}(\tau)t_{\mu\nu}(\tau)\nonumber \\
 & = & S_{NG}+S_{SS},\label{eq:defSss}\end{eqnarray}
 and consider the variation of each of these terms under SUSY1. The
general variation of a loop functional, $F[x(\tau)]$ , can be written
as \begin{equation}
\delta F=\oint\delta x_{\mu}dx_{\nu}\frac{\delta F}{\delta\sigma_{\mu\nu}}.\label{eq:loopDerivative}\end{equation}
 Using this the variation of $S_{NG}$ under (\ref{eq:wlineSUSY})
can be written as \begin{eqnarray}
\delta_{S1}S_{NG} & = & T_{0}\oint\delta x_{\mu}dx_{\nu}\frac{\delta S_{NG}}{\delta\sigma_{\mu\nu}}\nonumber \\
 & = & T_{0}\oint d\tau\dot{x}_{\nu}\epsilon\psi_{\mu}t_{\mu\nu}.\label{eq:deltaS1NG}\end{eqnarray}
 The variation of $S_{SS}$ under (\ref{eq:wlineSUSY}) is\begin{eqnarray}
\delta_{S1}S_{SS} & = & \delta_{S1}\left(-\frac{T_{0}}{2}\oint d\tau\psi_{\mu}\psi_{\nu}t_{\mu\nu}\right)\nonumber \\
 & = & -T_{0}\oint d\tau\dot{x}_{\nu}\epsilon\psi_{\mu}t_{\mu\nu}-\frac{T_{0}}{2}\oint d\tau\psi_{\mu}\psi_{\nu}\delta_{S1}t_{\mu\nu},\label{eq:deltaS1SS}\end{eqnarray}
 the first term in the above equation cancels with the variation of
$S_{NG}$ given by (\ref{eq:deltaS1NG}). Consider now the variation
of $t_{\mu\nu}$ under SUSY1,\begin{equation}
\delta_{S1}t_{\mu\nu}(x(\tau))=t_{\mu\nu}(x(\tau)+\epsilon\psi(\tau))-t_{\mu\nu}(x(\tau)),\label{eq:deltaS1tmunu}\end{equation}
 in the context of loop calculus this quantity can be represented
by a path derivative \citet{MIGDAL:1983aa,Makeenko:1999hq},\begin{eqnarray}
\delta_{S1}t_{\mu\nu}(x(\tau)) & = & \partial_{\lambda}^{x}t_{\mu\nu}\delta x_{\lambda},\nonumber \\
 & = & \epsilon\psi_{\lambda}\partial_{\lambda}^{x}t_{\mu\nu},\label{eq:deltaS1Pathderivative}\end{eqnarray}
 where $\partial_{\lambda}^{x}$ denotes the path-derivative at point
$x(\tau)$ and we have used (\ref{eq:wlineSUSY}). This allows us
to write the variation in the second term of (\ref{eq:deltaS1SS}),
using (\ref{eq:areaDerivativeNG}), as\begin{equation}
\oint d\tau\psi_{\mu}\psi_{\nu}\delta_{S1}t_{\mu\nu}=\oint d\tau\epsilon\psi_{\mu}\psi_{\nu}\psi_{\lambda}\partial_{\lambda}(\tau)\frac{\delta}{\delta\sigma_{\mu\nu}(\tau)}\left(\int d^{2}\sigma\sqrt{g}\right).\label{eq:deltaSss2}\end{equation}
 The area derivative satisfies a Bianchi identity\begin{equation}
\partial_{\lambda}(\tau)\frac{\delta}{\delta\sigma_{\mu\nu}(\tau)}+\partial_{\mu}(\tau)\frac{\delta}{{\delta\sigma}_{\nu\lambda}(\tau)}+\partial_{\nu}(\tau)\frac{\delta}{\delta\sigma_{\lambda\mu}(\tau)}=0,\label{eq:bianchiId}\end{equation}
 as a result the second term, (\ref{eq:deltaSss2}), in Eq. (\ref{eq:deltaS1SS})
vanishes and the action (\ref{eq:defSss}) is invariant under the
worldline SUSY transformation (\ref{eq:wlineSUSY}).

Having obtained the spin-string interaction, one would like to know
whether one can relate it to spontaneous breaking of chiral symmetry.
This can be done, at least formally, in the large $N$ limit using
Banks and Cashers relation \citet{Banks:1979yr} that expresses the
vacuum expectation value of chiral condensate, $V_{\chi}$ , in term
of the expectation value of a super Wilson loop

\begin{equation}
V_{\chi}=m\int_{0}^{\infty}dT\exp\left\{ -\frac{m^{2}}{2}T\right\} \int_{y,\psi}\exp\{-S_{0}\}<\W>_{YM},\label{eq:yPIforChi}\end{equation}
where the subscript $y,\psi$ under the integral represents a sum
over all closed paths of spin half particle whose length is $T$,
and $S_{0}$ is the action for a free spin-half particle,\begin{equation}
S_{0}=\int_{0}^{T}d\tau\Big\{\frac{\dot{x}^{2}}{2}+\frac{1}{2}\psi_{\mu}\dot{\psi_{\mu}}\Big\}\label{eq:freeSpinHalf}\end{equation}
 To check for spontaneous breaking of chiral symmetry one has to consider
the above expression in the limit $m\rightarrow0$, where $m$ is
the current quark mass. Using the string representation for the expectation
value of super Wilson loop (\ref{eq:defStringSWL}), we can write
chiral condensate as\begin{eqnarray}
V_{\chi} & = & \lim_{m\rightarrow0}m\int_{0}^{\infty}dT\exp\left\{ -\frac{m^{2}}{2}T\right\} \int_{y,\psi}\exp\{-S_{0}\}\nonumber \\
 & \times & \int DX\exp\left\{ -S_{SWL}[X,x(\tau),\psi(\tau)]\right\} .\label{eq:chiSWLstring}\end{eqnarray}
Unfortunately this is cumbersome and intractable as it involves sum
over infinite number of boundaries, and for each boundary one has
to sum over surfaces. But it does indicate the role of spin-string
interaction for describing the spontaneous breaking of chiral symmetry.

\section{Super Wilson Loop and the Heavy Quark Potential \label{sec:SWLstaticPot}}

\begin{figure}
\includegraphics[scale=0.45]{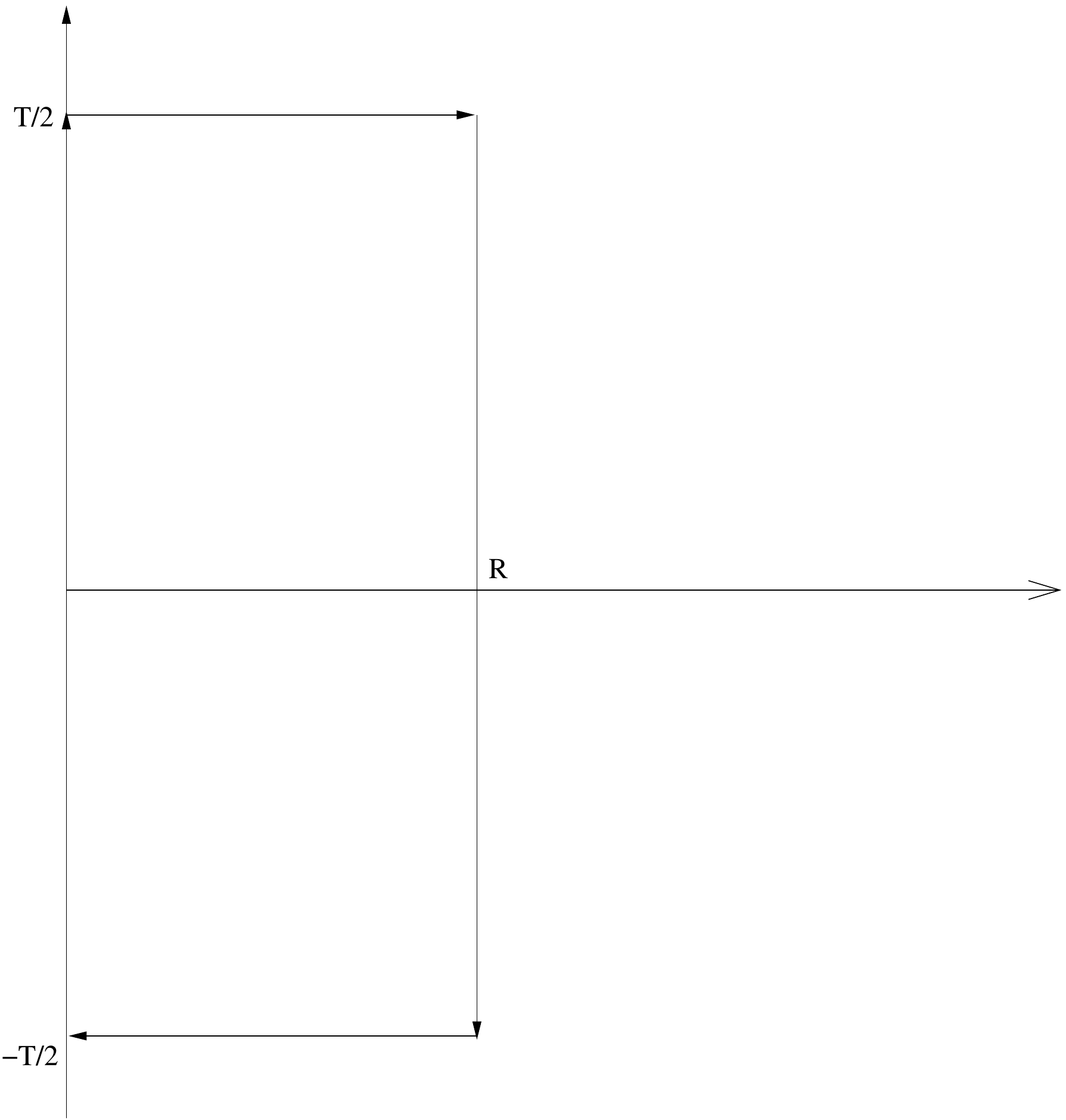}

\caption{Loop for calculating spin-dependent Heavy quark potential\label{fig:RectSWL}}

\end{figure}

The spin dependent corrections to the heavy quark potential can be
obtained from the expectation value of a rectangular super Wilson
loop \citet{Eichten:1980mw,Peskin:1983aa,Barchielli:1986zs,Gromes:1984ma,Brambilla:1993zw}.
For this purpose it will be more convenient to consider super Wilson
loop written in terms of the Dirac gamma matrices, Eq.(\ref{eq:WLspinHalf}),
and then consider the non-relativistic limit of the following amplitude\begin{equation}
Z_{q\bar{q}}=\int_{0}^{\infty}dT\int Dx\exp\left\{ -\int_{0}^{T}d\tau\frac{1}{2}(\dot{x}^{2}+m^{2})\right\} <\W>_{YM}.\label{eq:Zqqbar}\end{equation}
 In the non-relativistic limit the parameter $\tau$ is related to
the Euclidean time by\begin{equation}
\tau=\frac{x_{0}}{m}=\frac{t}{m},\label{eq:nrLimittau}\end{equation}
 where $m$ is the quark mass \citet{Peskin:1983aa}, and in the same
limit the super Wilson loop associated with a rectangular loop, Fig.(\ref{fig:RectSWL}),
is \begin{equation}
\W_{NR}[T,R]={\rm Tr}\hat{P}\left\{ i\oint dt(\dot{x}\cdot A)+\frac{1}{4m}\oint dt(\Sigma_{\mu\nu}F_{\mu\nu})\right\} ,\label{eq:rectSWL}\end{equation}
 where we have taken the limit $T\rightarrow\infty$ and ignored the
contribution from the short sides of the rectangular loop. The rectangular
loop can be thought of as being made of the worldline of a quark at
origin and a worldline of an antiquark located at a distance $R$
from it. According to our assumptions the expectation value of such
a super Wilson loop is given by \begin{equation}
<\W_{NR}>_{YM}=\int DX\exp\left\{ -S_{SWL}\right\} ,\label{eq:NrSWLexp}\end{equation}
 with the string action \begin{equation}
S_{SWL}[T,R]=T_{0}\int d^{2}\sigma\sqrt{g}+i\frac{T_{0}}{4m}\oint\Sigma_{\mu\nu}t_{\mu\nu}(x_{0})dt.\label{eq:SNRSWL}\end{equation}
 The expectation value of a rectangular super Wilson loop, in the
limit $T\rightarrow\infty$, can be expressed as\begin{equation}
<\W[T,R]>_{YM}=\exp\left\{ i\phi(T,R)\right\} \exp\left\{ -V(R)T\right\} ,\label{eq:swlTosdpot}\end{equation}
 where $\phi(T,R)$ is a phase factor which is a peculiarity of Euclidean
path-integrals for fermions, while $V(R)$ is the spin-dependent potential
between the quark and the antiquark separated by a distance $R$ (the
use of euclidean path integral to obtain spin-dependent potentials
is reviewed in \citet{Peskin:1983aa}.)

In extracting the spin-dependent potential, it is both suggestive
and convenient to write the spin-string interaction term as\begin{equation}
\Sigma_{\mu\nu}t_{\mu\nu}=\spin\cdot\B-\spin\cdot\E,\label{eq:tmuToFields}\end{equation}
 where the worldline quantities $\B$ and $\E$ are defined as\begin{eqnarray}
\mathcal{B}_{i} & = & {\frac{1}{2}\epsilon}_{ijk}t_{jk},\nonumber \\
\mathcal{E}_{i} & = & t_{oi},\label{eq:defBdefE}\end{eqnarray}
 and $\spin$ are the Pauli-spin matrices. In the non-relativistic
limit we can restrict to the upper two-components of the Dirac spinors.
For a rectangular super Wilson loop the {}``electric term'', $\spin\cdot\E$,
only contributes to a phase factor in Eq.(\ref{eq:swlTosdpot}) and
the spin-spin term arises from the {}``magnetic term'', $\spin\cdot\B$.
The string action for a rectangular super Wilson loop that contributes
to the heavy quark potential takes the form\begin{equation}
S_{SWL}[T,R]=T_{0}\int d^{2}\sigma\sqrt{g}+i\frac{T_{0}}{4m}\int dt^{+}\spin^{+}\cdot\B^{+}-i\frac{T_{0}}{4m}\int dt^{-}\spin^{-}\cdot\B^{-},\label{eq:rectSWLstring}\end{equation}
 where the superscripts $\pm$ denote the quark and the antiquark
S.

It will be convenient to introduce dimensionless coordinates,\begin{eqnarray}
M & = & \sqrt{T_{0}},\nonumber \\
Y(\sigma_{0},\sigma_{1}) & = & MX(\sigma_{0},\sigma_{1}),\label{eq:dimlessCoord}\end{eqnarray}
 and the small transverse fluctuations of the minimal surface \begin{equation}
\phivector=(Y_{2},Y_{3})=(\phi_{y},\phi_{z}),\label{eq:phiVector}\end{equation}
 can be parametrized using\begin{equation}
\sigma_{0}=Y_{0}=\bar{t};\,\sigma_{1}=Y_{1}=\bar{r}.\label{eq:dimParameter}\end{equation}
 In terms of these dimensionless variables the action for the rectangular
super Wilson loop is \begin{equation}
S_{SWL}[T,R]=\int d\tbar\int d\rbar\sqrt{g}+i\frac{M}{4m}\int d\tbar^{+}\spin^{+}\cdot\B^{+}-i\frac{M}{4m}\int d\tbar^{-}\spin^{-}\cdot\B^{-}.\label{eq:SdimLess}\end{equation}
 The appropriate boundary conditions for a rectangular super Wilson
loop are\begin{equation}
\partial_{\tbar}\phivector(\tbar,0)=\partial_{\tbar}\phivector(\tbar,\bar{R})=0,\label{eq:rectBC}\end{equation}
 with these boundary conditions, and using \begin{equation}
t_{ij}=\frac{1}{\sqrt{g}}(\dot{\phi_{i}}\phi_{j}^{\prime}-\phi_{i}^{\prime}\dot{\phi_{j}}),\label{eq:tijForRectSWL}\end{equation}
 we immediately see that\begin{equation}
\B^{+}=\B^{-}=0\label{eq:stringB}\end{equation}
 and therefore there is no contribution from the spin-string interaction
to the heavy quark-antiquark potential, and in particular there is
no spin-spin dependent term in the heavy quark potential.

Absence of a spin-spin term in the heavy quark potential seems to
be consistent with the experimental results and lattice simulations.
These results suggest that quarks see purely chromoelectric field
in their rest frame \citet{Buchmuller:1981fr}, and is the starting
point for introducing spin degrees of freedom in open string models
of mesons in \citet{Allen:2003kd}.

\section{Spin-String Interaction in the Flux-tube Model\label{sec:fluxTubeStaticPot}}

The absence of a spin-spin dependent correction to the heavy quark
potential is perhaps surprising, for there is an argument due to Kogut
and Parisi \citet{Kogut:1981aa}, in the context of the flux-tube
model of confinement, for the existence of a long range spin-spin
dependent term in the heavy quark potential. They argue, using the
language of $U(1)$ gauge theory, that the zero-point fluctuations
of the flux-tube creates a time-dependent electric flux lines which
in turn produces a magnetic field. This magnetic field interacts with
the spin of the quark and the antiquark, leading to a spin-spin interaction
term in the heavy quark potential. The argument in the previous section
implies that the {}``magnetic'' field vanishes on the quark worldline
for a static quark or antiquark, but the argument is for the {}``magnetic''
field produced by a string with no intrinsic thickness and could get
modified for a flux-tube which has finite intrinsic thickness. One
possible way of taking into account the intrinsic thickness of the
flux-tube, while still retaining the effective string description,
is to evaluate the {}``magnetic'' field $t_{ij}$, not at the boundary,
but to average it over a longitudinal distance of the order of the
intrinsic thickness of the string, $r_{T}$,\begin{eqnarray}
\bar{t}_{ij}(t,x^{+}) & = & \frac{1}{r_{T}}\int_{0}^{r_{T}}drt_{ij}(t,r)\nonumber \\
 & = & \frac{1}{r_{T}}\int_{0}^{r_{T}}dr\left(t_{ij}(t,x^{+})+r\partial_{r}t_{ij}(t,x^{+})\right)\nonumber \\
 & = & \frac{r_{T}}{2}(\partial_{r}t_{ij})_{x^{+}}.\label{eq:avgTij}\end{eqnarray}
 Evaluating $\bar{t}_{ij}$ for small transverse fluctuations, $\phi_{i}\ll1$,
for which\begin{equation}
\sqrt{g}=1+\frac{1}{2}(\partial_{\tbar}\phivector^{2}+\partial_{\rbar}\phivector^{2}),\label{eq:gaussG}\end{equation}
 and keeping only the leading terms in $\phivector$, we find a non-vanishing
spin-string interaction\begin{equation}
\spin\cdot\bar{\B}(x^{\pm})=\frac{1}{M_{g}}\spin\cdot(0,\partial_{r}\dot{\phi}_{z}(x^{\pm}),-\partial_{r}\dot{\phi}_{y}(x^{\pm})).\label{eq:avgSpinString}\end{equation}
 where $M_{g}^{-1}=r_{T}/2$ is some measure of the intrinsic thickness
of the flux-tube. Apart from the factor of $M_{g}^{-1}$, this is
precisely the interaction assumed by Kogut and Parisi in Ref.\citet{Kogut:1981aa}.
Using this spin-string interaction, the action for small transverse
fluctuations about the minimal surface binding the rectangular loop
is\begin{equation}
S_{sd}=\bar{R}\bar{T}+\frac{1}{2}\int d\bar{t}d\bar{r}(\partial_{\bar{t}}\dot{\phivector}^{2}+\partial_{\bar{r}}\dot{\phivector}^{2})+i\alpha_{ss}\int d\bar{t}^{+}\spin\cdot{\bf b}-i\alpha_{ss}\int d\bar{t}^{-}\spin\cdot{\bf b},\label{eq:pertSpinString}\end{equation}
 where the spin-string coupling constant is\begin{equation}
\alpha_{ss}=\frac{T_{0}}{4mM_{g}}\label{eq:effStringSpinCouplingConstant}\end{equation}
 and the dimensionless {}``magnetic'' field is \begin{equation}
{\bf b}=(0,\partial_{r}\dot{\phi}_{z}(x^{\pm}),-\partial_{r}\dot{\phi}_{y}(x^{\pm})).\label{eq:pertBfield}\end{equation}
 The expectation value of the rectangular super Wilson loop then

\begin{eqnarray}
< & \W_{NR}>_{YM} & =\exp\left\{ -\bar{R}\bar{T}\right\} Z_{RT}\times\nonumber \\
 &  & <\exp\left\{ -i\alpha_{ss}(\int d\bar{t}^{+}\spin\cdot{\bf b}-\int d\bar{t}^{-}\spin\cdot{\bf b})\right\} >_{\phi}\label{eq:perZspinString}\end{eqnarray}

where the average over the string fluctuations $\phivector$ is given
by\begin{equation}
Z_{RT}=\int_{\phivector}\exp\left\{ -\frac{1}{2}\int d\bar{t}d\bar{r}(\partial_{\bar{t}}\dot{\phivector}^{2}+\partial_{\bar{r}}\dot{\phivector}^{2})\right\} .\label{eq:Zgaussian}\end{equation}
 If we set $\alpha_{ss}$ to zero then the super Wilson loop reduces
to the Wilson loop and we recover the linear potential along with
the Lüscher term. The effect of the spin-string interaction ${\bf b}\cdot\spin$
can be evaluated in perturbation theory in a manner identical to that
of Ref.\citet{Kogut:1981aa} and the first non-vanishing term appears
in the fourth-order in $\alpha_{ss}$ and gives rises to the\begin{equation}
V_{ss}=\frac{T_{0}^{2}}{(mM_{g})^{4}}\frac{\spin^{+}\cdot\hat{R}\,\spin^{-}\cdot\hat{R}}{R^{5}},\label{eq:KPspin-spin}\end{equation}
 where $\hat{R}$ is a unit vector pointing from the quark to the
antiquark and dimensionless numerical factors have been absorbed in
$M_{g}$ whose inverse we have taken as a measure of the thickness
of the flux-tube. In the limit $M_{g}^{-1}\rightarrow0$, which corresponds
to a flux-tube with no intrinsic thickness, $V_{ss}$ vanishes and
there is no spin-spin correction to the heavy quark potential due
to spin-string interaction. 

It is worth emphasizing that our calculation is entirely within an
effective string description. We have only modified the spin-string
interaction, in Eq. (\ref{eq:avgTij}), by averaging it along the
string rather than restricting it to the boundary. Thus, the dynamics
is that of a string with no intrinsic thickness but with a modified
spin-string interaction. It is because of this and particularly because
of the ground state fluctuations of the string, that we obtain a long
range spin-spin interaction (\ref{eq:KPspin-spin}) and this is also
the reason for the vanishing of the second order term in spin-spin
interaction which is proportional to $1/m^{2}$ (see \citet{Kogut:1981aa}
for details.) Our effective string model, by definition, does not
include the short-range correlations which are responsible for the
formation of the flux tube and which give rise to exponentially decaying
spin-spin interaction of the order $1/m^{2}$ with a decay length
proportional to the the intrinsic thickness of the flux tube\citet{Brambilla:1997aa,Baker:1996mk}.
We comment on a possible way of exploring the relationship between
a fundamental string and a flux-tube in the next section.

\section{Conclusions\label{sec:Conclusions}}

In a string description of QCD it is important to find out the nature
of the spin-string interaction, as it can illuminate both the spin-dependent
corrections to the heavy quark potential and with in the context of
a fundamental string description it may also help us in understanding
the existence of a massless pion in chiral limit and more generally
understand the pion-rho mass difference. The approach we have take
to analyse this question is to write the expectation value of a super
Wilson loop as a sum over surfaces whose boundary is the given loop.
Each surface appearing in the sum can be interpreted as a world-sheet
of an open string that terminates on a worldline of a spin-half particle,
the quark in our case. The action appearing in the string representation
then naturally includes the spin-string interaction.

In order to obtain a string representation for the expectation value
of the super Wilson loop, we used the fact that the super Wilson loop
is related to the Wilson loop via the area derivative of a loop. Then
we assumed that the expectation value of the Wilson loop has a string
representation with the string action for large loops being the Nambu-Goto
action. The resulting string action for the super Wilson loop is the
Nambu-Goto action with an additional boundary term that incorporates
the interaction between the spin degrees of freedom and the string
degrees of freedom. The super Wilson loop is invariant under a worldline
SUSY, the string action that we have obtained has the desired property
that it too is invariant under this symmetry. An important question
that we have not discussed in the present investigation is the relationship
between the string representation of the super Wilson loop and the
string model of mesons which include spin quantum number. A formal
string representation for the meson propagator can of course be written
in terms of the expectation value of super Wilson loop, in a manner
very similar to the expression for chiral condensate (\ref{eq:chiSWLstring}),
but it does not provide a direct string representation for the mesons. 

One can extract the spin-dependent potential from the expectation
value of a rectangular super Wilson loop. We found that the spin-string
interaction does not contribute to heavy quark potential. But if we
try and incorporate the effect of the finite intrinsic thickness of
the flux-tube by averaging the spin-string interaction over a longitudinal
distance of the order of thickness of the flux-tube, then we do obtain
a spin-spin term in the heavy quark potential. The form of the resulting
term is precisely the one suggested by Kogut and Paris based on the
fluctuation of the electric field lines forming a flux-tube \citet{Kogut:1981aa}.
The spin-spin interaction that we obtained depends, in addition to
the mass of the quark, on the square of the string tension and on
the intrinsic thickness of the flux-tube.

In the context of an effective string description of QCD the idea
of an intrinsic thickness of a flux-tube remains heuristic. It is
quite plausible that the flux-tube in QCD has an intrinsic thickness,
but in the absence of our understanding of the physics behind confinement
we cannot identify an operator whose expectation value would give
the thickness of the flux-tube. AdS/CFT correspondence could perhaps
illuminate this issue. In ref. \citet{Polchinski:2001ju} the authors
have argued, using AdS/CFT correspondence, that while the hadrons
are represented by an ideal fundamental string with no intrinsic thickness
in the the bulk of the 5-dimensional AdS space, but their holographic
projection on to the 4-dimensional boundary theory do have a finite
intrinsic thickness. Therefor it would be very interesting and useful
to try and obtain a string representation for the expectation value
of super Wilson loop using AdS/CFT correspondence and to see if there
are any spin-spin terms in the heavy quark potential so obtained.

\section*{Acknowledgments}

I would like to thank Gunnar Bali for his very useful comments on
the preliminary version of this work. I have also benefited from the
comments of Theodore Allen and would like to thank him for that. This
work was started while I was visiting the Institute for Mathematical
Sciences, Chennai. I would like to thank Hari Dass for inviting me
to visit IMSC and for various useful discussions that I had with him
during my visit. At IMSC, I would also like to thank Rahul Basu for
greatly facilitating my visit. This paper is dedicated to the memory
of Prof. S. C. Bhargava and Prof. K. Swaminathan, both of whom taught
at the physics department of St. Stephen's college, Delhi Univerity.

%\bibliographystyle{/Users/vikram/Documents/Physics/BibTeX/revtex}
%\addcontentsline{toc}{section}{\refname}\bibliography{/Users/vikram/Documents/Physics/BibTeX/vik}

\end{document}